**Title:** Steady-state nonlinear optical response of excitons in monolayer MoSe$_2$

**Author Names:** Muhed S. Rana[1], Joshua R. Hendrickson[2], Christopher E. Stevens[2-3], Michael R. Koehler[4], David G. Mandrus[5-7], Takashi Taniguchi[8], Kenji Watanabe[9], Nai H. Kwong[10], Rolf Binder[1,10]* and John R. Schaibley[1]*

**Author Addresses:**
[1]Department of Physics, University of Arizona, Tucson, Arizona 85721, USA
[2]Air Force Research Laboratory, Sensors Directorate, Wright-Patterson Air Force Base, Ohio, 45433, USA
[3]KBR, Inc., Beavercreek, Ohio 45431, USA
[4]JIAM Diffraction Facility, Joint Institute for Advanced Materials, University of Tennessee, Knoxville, Tennessee 37920, USA
[5]Department of Materials Science and Engineering, University of Tennessee, Knoxville, Tennessee 37996, USA
[6]Materials Science and Technology Division, Oak Ridge National Laboratory, Oak Ridge, Tennessee 37831, USA
[7]Department of Physics and Astronomy, University of Tennessee, Knoxville, Tennessee 37996, USA
[8]International Center for Materials Nanoarchitectonics, National Institute for Materials Science, 1-1 Namiki, Tsukuba 305-0044, Japan
[9]Research Center for Functional Materials, National Institute for Materials Science, 1-1 Namiki, Tsukuba 305-0044, Japan
[10]James C. Wyant College of Optical Sciences, University of Arizona, Tucson, Arizona, 85721, USA

**Corresponding Authors:** John Schaibley, johnschaibley@email.arizona.edu
Rolf Binder, binder@optics.arizona.edu

**Abstract:**

Monolayer transition metal dichalcogenide (TMD) semiconductors such as $MoSe_2$ host strongly bound excitons which are known to exhibit a strong resonant third-order nonlinear response. Although there have been numerous studies of the ultrafast nonlinear response of monolayer TMDs, a study of the steady-state nonlinear response is lacking. We report a comprehensive study of the steady-state two-color nonlinear response of excitons in hBN-encapsulated monolayer $MoSe_2$ at 7 K. We observe differential transmission (DT) signals associated with the neutral and charged exciton species, which is strongly dependent on the polarization of the pump and probe. Our results are compared to a theoretical model based on a T-matrix formulation for exciton-exciton, exciton-trion, and trion-trion correlations. The parameters are chosen such that the theory accurately reproduces the experimental DT spectrum, which is found to be dominated by two-exciton correlations without strong biexciton binding, exciton-trion attractive interactions, and strong spin mixing through incoherent relaxation.

**Main Text:**

**Introduction:**

The linear and nonlinear optical response of excitons in monolayer transition metal dichalcogenide (TMD) semiconductors has been the subject of intense research over the past decade for their potential in various optoelectronic applications, such as the creation of optical switches, buffers, and modulators that mediate interactions between two beams[1–5]. In this work, we study monolayer $MoSe_2$, which is known to exhibit nearly homogeneous ~2-3 meV exciton linewidths when encapsulated with hexagonal boron nitride (hBN)[6–8]. The low energy "A" exciton of monolayer $MoSe_2$ has a binding energy of ~500 meV[9–11] and dominates the optical response. This resonance consists of two direct optical transitions that occur at the K and K' momentum space valleys on the edge of the hexagonal Brillouin zone and can be addressed optically using the valley dependent optical selection rule[10,12,13]. In addition to neutral excitons, it is known that monolayer TMDs are often intrinsically doped with electrons and can therefore host charged exciton species, referred to as trions or attractive polarons, with a binding energy of ~25-30 meV[9,10]. Previous studies have used ultrafast pump-probe spectroscopy[14] to probe the $MoSe_2$ exciton linewidth[8], trion formation dynamics[15,16], biexciton formation[16–18], and an Autler-Townes type exciton splitting[19].

In this work, we use low temperature (7 K) steady-state non-degenerate differential transmission spectroscopy to probe the valley dependent exciton interactions in high quality monolayer $MoSe_2$ encapsulated in hBN. Steady-state non-degenerate DT spectroscopy is a two-color technique where the continuous wave (CW) pump and probe lasers are scanned in photon energy independently, allowing for a measurement of the two-dimensional pump-probe nonlinear response. The spectral profiles of the DT spectra reveal information about the physical origin of the nonlinear response, which can include effects such as saturation, resonance shifts, and excitation induced dephasing[20–22]. A similar technique was previously used to probe population pulsation of neutral excitons in monolayer $MoSe_2$[23] but did not report the two-dimensional differential transmission spectrum that probes the nature of the interactions between excitons and trions. By comparing our experimental results to a recently developed T-Matrix model for the nonlinear response in the vicinity of the exciton resonance[24], we are able to explain the complex differential transmission lineshapes in terms of two-exciton states without a strongly bound biexciton, but with exciton trion correlations and spin/valley mixing through incoherent relaxation.

**Experimental Results:**

In order to probe the steady-state nonlinear response, we performed two-color DT/T measurements where we varied both pump and probe photon energies, depicted in Figure 1a. In the experiments, we measured the transmitted probe laser, and the (cross-polarized) pump was rejected with a polarizer. Figure 1b shows an example linear transmission spectrum (blue) and photoluminescence (PL) spectrum (red). The neutral exciton PL spectrum has a center energy of 1.652 eV and a full width half maximum linewidth of 3.9 meV. The DT signal is normalized by the linear transmission (T) to give the dimensionless DT/T nonlinear response. See Supplementary Note 1 for additional details of experimental methods. Figure 2a(b) shows a map of the cross linearly (circularly) polarized DT/T response as a function of pump and probe photon energies. For both cross-linearly and cross-circularly polarized pump and probe, the largest DT/T signal occurs when the pump and probe are resonant with the neutral exciton at ~1.652 eV. We note that if this line shape is interpreted in terms of a pump induced change to a Lorentzian transmission profile, the dominant nonlinear line shape corresponds to a pump induced red shift[25] (see Supplementary Figure 1). The cross-linearly polarized DT/T response also shows significant signal when pumping at 1.652 eV and probing at 1.626 eV, i.e., near MoSe$_2$ trion (charged exciton) resonance[10]. Line cuts of the DT/T spectra at pump photon energy of 1.652 eV are shown in Figure 2c-d. We also observe a significant DT/T response when pumping at 1.626 eV (trion) and probing near 1.652 eV (exciton). We again note that while the exact DT/T lineshapes for these exciton-trion nonlinear responses depend on the exact pump energy, the basic (negative to positive) profile is consistent with a pump induced red shift interpretation.

Figure 2b shows the cross-circularly polarized pump probe DT/T map. In addition to the pump-exciton probe-exciton signal at 1.652 eV, there is a negative DT/T response when probing near 1.626 eV, which appears independently of the pump photon energy above 1.626 eV. There is also a negative to weakly positive DT/T line shape when pumping at 1.626 eV (trion) and probing at 1.652 eV (exciton). The biexciton resonance reported in previous work[16–18] (occurring at ~20 meV below the neutral exciton) is absent and will be discussed in detail below. One surprising feature of the cross-circularly polarized response is the small positive feature that appears when pumping at 1.652 eV (exciton) and probing exactly at 1.626 eV (near the trion). The blue curves of Figure 2c-d show the cross-circularly polarized spectrum when pumping at 1.652 eV. Supplementary Figure 2a shows the high-resolution DT/T map when pumping the exciton and probing the trion. Supplementary Figure 2b shows the power dependence of the DT/T spectrum for the pump energy fixed at 1.652 eV. The DT/T signal is negative at low power, but the narrow positive feature emerges at ~ 50 µW pump power with a probe energy of 1.628 eV (Supplementary Figure 2b).

**Theoretical Results and Discussion:**

In the following, we briefly summarize the key features of the theoretical model used to analyze the experimental data. A detailed description of the model is given in reference 24. We use a similar approach to Takayama et al.[26] originally developed for the nonlinear optical response in the coherent third-order regime, where the differential absorption and DT are given in terms of products of linear susceptibilities and the T-matrix describing non-perturbative 2-particle scattering. In our model (Figure 3), the particles are either bright excitons or trions (both types being treated as point particles), and the free 2-particle propagator depends on both intravalley and intervalley long-range electron-hole (e-h) exchange interactions, with the former only changing the shape of the dispersion relations (the particle's center of mass energy) and the latter coupling the spin and valley states. The effect of the short-range e-h exchange is included in the exciton energy at zero momentum[27], and dark (spin forbidden) intervalley excitons are omitted in the model. The coupling of the spin states due to intervalley exchange is associated with a change of the relative orbital angular momentum (OAM) of the two particles (spin-orbit coupling). An analytic expression of the e-h exchange interaction can be found in Yu et al[28]. Additional details of the theoretical model can be found in the Supplementary Notes 2-4 and Supplementary Figures 3-5.

The interaction strength between each pair of particles (i.e., exciton-exciton, exciton-trion, and trion-trion) are adjustable parameters. We are using a separable interaction (compare Haug et al[29]), which we choose to be independent of the wavevector (relative momentum of the two particles) up to a cut-off wavevector, above which it is zero. Since the relative momentum is a vector in two dimensions, we expand the potential in an OAM series. For each combination of the particle species, spin, and relative OAM, we have one parameter characterizing the strength and sign of the interaction (negative for attractive, positive for repulsive).

We optimized a single set of interaction parameters to obtain theoretical results that reproduce accurately all the essential features seen in the experimental two-color DT map. The numerical results of this model are plotted in Figure 4a-b. Focusing on the line cuts with the pump centered at the exciton (Figure 4c), we find that the DT in the vicinity of the exciton is negative-positive-(weakly) negative, in good agreement with the experimental results in Figure 2c. Apart from the weakly negative feature at the high energy side of the exciton, this signifies a red shift of the exciton. As discussed in more detail in the Supplementary Notes 3-4 and Supplementary Figures 6-7, exciton-exciton and exciton-trion complexes exhibit scattering resonances (or correlated states) in the vicinity of their respective 2-particle continuum edges rather than forming discrete bound states, such as a biexciton. This is a consequence of the long-range e-h exchange interaction, which modifies the exciton dispersion relation. With that, we find that the experimentally observed nonlinear spectra are consistent with the interpretation of resonance enhancement at the trion and exciton resonance, respectively, together with a correlated exciton-trion state. Our model also includes a small amount of pump-induced free carrier generation (order of $10^8$ cm$^{-2}$ when pump is at exciton resonance, and smaller values otherwise), which enhances the negative DT/T signal in the cross-circularly polarized configuring when probing the trion. Our model also reproduces the red-shift-like response when pumping and probing the neutral exciton in the cross-linearly polarized configuration. This is achieved in a phenomenological fashion by using an effective attractive interaction in the co-circularly polarized T-matrix channel (which largely determines the shape of the DT in the cross-linear configuration[26]) (see Supplementary Figure 4). While individual co-circularly polarized exciton pairs have a repulsive interaction, in a CW experiment, pumping, say, spin-up excitons will indirectly lead to spin down excitons through incoherent spin/valley relaxation, thus enabling spin-up probe excitons to interact attractively with the indirectly created excitons of opposite spin, which can lead to a response that dominates over that from the repulsive interaction in the same valley/spin state. This spin mixing would not be expected to be present in ultrafast spectroscopy probing the coherent nonlinear response[30]. Since incoherent spin/valley relaxation tends to equalize the exciton populations in both spin/valley states, it cannot explain the absence of a biexciton in the cross-circular configuration, because that would require all pumped spin-up excitons to relax into spin-down states, so that there would be no cross-circular signal left when the spin-down state is probed.

We see in Figures 2a,b and 4a,b that the DT signal occurs mainly on two vertical lines of fixed pump frequency $\omega_p$, which is indicative of response due to resonance enhancement of the exciton and trion resonance that are seen also in linear spectroscopy. The experimental data show no visible signal at the two-photon resonance frequency where pump and probe frequencies add up to the biexciton energy, $\hbar\omega_p + \hbar\omega_{pr} = \varepsilon_{biexciton}$, which would provide a stringent proof for a biexciton resonance in our own and other experiments. However, the absence of such a signal can be due to insufficient signal strength and cannot by itself be used as a proof for the absence of the biexciton. We emphasize that the lack of a strongly bound biexciton resonance is consistent in our theoretical and experimental cross-circularly polarized results. As described in reference 24, this occurs from a reduction of the biexciton binding energy due to the strong e-h exchange interaction. Our result contrasts with past works which clearly observe a biexciton resonance in cross-circularly polarized nonlinear spectroscopy of monolayer MoSe$_2$, ~20 meV below the neutral exciton[15–17]. There are several possibilities that may explain the apparent contradiction, including variable sample quality, intrinsic doping level, peak excitation intensities. In our experiment, the spectrally integrated PL intensity from the exciton is greater than three times that of the trion, and the exciton linewidth (full width half maximum) is 3.9 meV, indicating potentially higher sample quality relative to past studies[15,16]. We speculate that in certain experimental conditions, competing relaxation processes may suppress the effect of the e-h exchange interaction leading to the formation of a strongly bound biexciton state in monolayer MoSe$_2$. We note that reports of biexciton PL in monolayer WSe$_2$ may be related the opposite conduction band splitting and presence of the reservoir of dark (spin forbidden) excitons[31,32].

**Summary:**

In this work, we have measured the steady-state nonlinear response of excitons and trions in ultra-high quality monolayer MoSe$_2$ and developed a theoretical model that reproduces the complex steady-state pump-probe DT response. Our results in the cross-circular configuration are consistent with a theory in which long-range e-h exchange strongly reduces the binding energy of two-exciton and exciton-trion complexes. In the cross-linear configuration, they indicate strong incoherent spin/valley relaxation, which would not be expected[30] in the ultrafast nonlinear response regime. Our results motivate further investigation of the two-color nonlinear response of the other 2D semiconductors hosting disparate exciton species as well as the doping dependence of monolayer MoSe$_2$, which has recently been shown to host optical signatures of Wigner-crystals for finite doping[33,34]. The simplicity of the theoretical model developed here and in reference 24 is designed to make it transparent and yield insight into basic aspects of the origins of the nonlinear exciton response. Future validation using theories starting from a fermionic theory of electrons and holes along the lines of Refs.[18,35], extended to account for all relevant incoherent relaxation and scattering processes involving excitons, trions and electrons, especially those that affect the e-h exchange interaction, account for the effects of two-particle continuum states on the DT lineshape, and account for dark-dark and mixed dark-bright biexcitons including all relevant incoherent scattering processes, are expected to shed further light on the nonlinear excitonic response in CW spectroscopy and to help apply the effects in novel optoelectronic technologies.

**Acknowledgments:**

**General:** JRS acknowledges useful discussions with Wang Yao, Hongyi Yu and Xiaodong Xu.

**Funding:** JRS acknowledges support from AFOSR (Grant Nos. FA9550-17-1-0215, FA9550-18-1-0390, and FA9550-20-1-0217 and the U.S. National Science Foundation (Grant Nos. ECCS-1708562, DMR-1838378 and DMR- 2054572). RB acknowledges support from the U.S. National Science Foundation (NSF), Grant No. DMR-1839570. DGM acknowledges support from the Gordon and Betty Moore Foundation's Epics Initiative, Grant GBMF9069. K.W. and T.T. acknowledge support from the Elemental Strategy Initiative conducted by the MEXT, Japan (Grant No. JPMXP0112101001) and JSPS KAKENHI (Grant Nos. 19H05790 and JP20H00354). JRH acknowledges support from the Air Force Office of Scientific Research (Program Manager Dr. Gernot Pomrenke) under award number FA9550-20RYCOR059.

**Author Contributions:**

JRS conceived and supervised the experimental aspects of the project. MSR fabricated the structures, and performed the experiments, and analyzed the data with input from JRS, JH and CES. MRK and DGM provided and characterized the bulk $MoSe_2$ crystals. TT and KW provided the hBN crystals. NHK and RB developed the theoretical model and theoretically analyzed the experimental data. MSR, JRS, and RB wrote the paper. All authors discussed the results.

**Competing Interests:**

The authors declare no competing interests.

**References:**

1. Ferrari, A. C. *et al.* Science and technology roadmap for graphene, related two-dimensional crystals, and hybrid systems. *Nanoscale* **7**, 4598–4810 (2015).

2. Wang, Q. H., Kalantar-Zadeh, K., Kis, A., Coleman, J. N. & Strano, M. S. Electronics and optoelectronics of two-dimensional transition metal dichalcogenides. *Nat. Nanotechnol.* **7**, 699–712 (2012).

3. Sun, Z., Martinez, A. & Wang, F. Optical modulators with 2D layered materials. *Nat. Photonics* **10**, 227–238 (2016).

4. Lee, E., Yoon, Y. S. & Kim, D.-J. Two-Dimensional Transition Metal Dichalcogenides and Metal Oxide Hybrids for Gas Sensing. *ACS Sens.* **3**, 2045–2060 (2018).

5. Chen, H., Wang, C., Ouyang, H., Song, Y. & Jiang, T. All-optical modulation with 2D layered materials: status and prospects. *Nanophotonics* **9**, 2107–2124 (2020).

6. Scuri, G. *et al.* Large Excitonic Reflectivity of Monolayer $MoSe_2$ Encapsulated in Hexagonal Boron Nitride. *Phys. Rev. Lett.* **120**, 037402 (2018).

7. Back, P., Zeytinoglu, S., Ijaz, A., Kroner, M. & Imamoğlu, A. Realization of an Electrically Tunable Narrow-Bandwidth Atomically Thin Mirror Using Monolayer $MoSe_2$. *Phys. Rev. Lett.* **120**, 037401 (2018).

8. Martin, E. W. *et al.* Encapsulation Narrows and Preserves the Excitonic Homogeneous Linewidth of Exfoliated Monolayer MoSe$_2$. *Phys. Rev. Appl.* **14**, 021002 (2020).

9. Wang, G. *et al.* Exciton states in monolayer MoSe$_2$ : impact on interband transitions. *2D Mater.* **2**, 045005 (2015).

10. Ross, J. S. *et al.* Electrical control of neutral and charged excitons in a monolayer semiconductor. *Nat. Commun.* **4**, 1474 (2013).

11. Ugeda, M. M. *et al.* Giant bandgap renormalization and excitonic effects in a monolayer transition metal dichalcogenide semiconductor. *Nat. Mater.* **13**, 1091–1095 (2014).

12. Mak, K. F., He, K., Shan, J. & Heinz, T. F. Control of valley polarization in monolayer MoS$_2$ by optical helicity. *Nat. Nanotechnol.* **7**, 494–498 (2012).

13. Xiao, D., Liu, G.-B., Feng, W., Xu, X. & Yao, W. Coupled Spin and Valley Physics in Monolayers of MoS$_2$ and Other Group-VI Dichalcogenides. *Phys. Rev. Lett.* **108**, 196802 (2012).

14. Selig, M. *et al.* Ultrafast dynamics in monolayer transition metal dichalcogenides: Interplay of dark excitons, phonons, and intervalley exchange. *Phys. Rev. Res.* **1**, 022007 (2019).

15. Hao, K. *et al.* Coherent and Incoherent Coupling Dynamics between Neutral and Charged Excitons in Monolayer MoSe$_2$. *Nano Lett.* **16**, 5109–5113 (2016).

16. Hao, K. *et al.* Neutral and charged inter-valley biexcitons in monolayer MoSe$_2$. *Nat. Commun.* **8**, 15552 (2017).

17. Yong, C.-K. *et al.* Biexcitonic optical Stark effects in monolayer molybdenum diselenide. *Nat. Phys.* **14**, 1092–1096 (2018).

18. Steinhoff, A. *et al.* Biexciton fine structure in monolayer transition metal dichalcogenides. *Nat. Phys.* **14**, 1199–1204 (2018).

19. Yong, C.-K. *et al.* Valley-dependent exciton fine structure and Autler–Townes doublets from Berry phases in monolayer MoSe$_2$. *Nat. Mater.* **18**, 1065–1070 (2019).

20. Jakubczyk, T. *et al.* Radiatively Limited Dephasing and Exciton Dynamics in MoSe$_2$ Monolayers Revealed with Four-Wave Mixing Microscopy. *Nano Lett.* **16**, 5333–5339 (2016).

21. Titze, M., Li, B., Zhang, X., Ajayan, P. M. & Li, H. Intrinsic coherence time of trions in monolayer MoSe$_2$ measured via two-dimensional coherent spectroscopy. *Phys. Rev. Mater.* **2**, 054001 (2018).

22. Katsch, F., Selig, M. & Knorr, A. Exciton-Scattering-Induced Dephasing in Two-Dimensional Semiconductors. *Phys. Rev. Lett.* **124**, 257402 (2020).

23. Schaibley, J. R. *et al.* Population Pulsation Resonances of Excitons in Monolayer MoSe$_2$ with Sub-μeV Linewidths. *Phys. Rev. Lett.* **114**, 137402 (2015).

24. Kwong, N.-H., Schaibley, J. R. & Binder, R. Effect of intravalley and intervalley electron-hole exchange on the nonlinear optical response of monolayer MoSe$_2$. arXiv preprint arXiv:2108.00091 (2021).

25. Chernikov, A., Ruppert, C., Hill, H. M., Rigosi, A. F. & Heinz, T. F. Population inversion and giant bandgap renormalization in atomically thin WS2 layers. *Nat. Photonics* **9**, 466–470 (2015).

26. Takayama, R., Kwong, N.-H., Rumyantsev, I., Kuwata-Gonokami, M. & Binder, R. Material and light-pulse parameter dependence of the nonlinear optical susceptibilities in the coherent $\chi^{(3)}$ regime in semiconductor quantum wells. *JOSA B* **21**, 2164–2174 (2004).

27. Qiu, D. Y., Cao, T. & Louie, S. G. Nonanalyticity, Valley Quantum Phases, and Lightlike Exciton Dispersion in Monolayer Transition Metal Dichalcogenides: Theory and First-Principles Calculations. *Phys. Rev. Lett.* **115**, 176801 (2015).

28. Yu, T. & Wu, M. W. Valley depolarization due to intervalley and intravalley electron-hole exchange interactions in monolayer $MoS_2$. *Phys. Rev. B* **89**, 205303 (2014).

29. Haug, H. & Schmitt-Rink, S. Electron theory of the optical properties of laser-excited semiconductors. *Prog. Quantum Electron.* **9**, 3–100 (1984).

30. Rodek, A. *et al.* Local field effects in ultrafast light–matter interaction measured by pump-probe spectroscopy of monolayer $MoSe_2$. *Nanophotonics* **0**, 20210194 (2021).

31. Zhang, M., Fu, J., Dias, A. C. & Qu, F. Optically dark excitonic states mediated exciton and biexciton valley dynamics in monolayer $WSe_2$. *J. Phys. Condens. Matter* **30**, 265502 (2018).

32. Bragança, H., Riche, F., Qu, F., Lopez-Richard, V. & Marques, G. E. Dark-exciton valley dynamics in transition metal dichalcogenide alloy monolayers. *Sci. Rep.* **9**, 4575 (2019).

33. Smoleński, T. *et al.* Signatures of Wigner crystal of electrons in a monolayer semiconductor. *Nature* **595**, 53–57 (2021).

34. Zhou, Y. *et al.* Bilayer Wigner crystals in a transition metal dichalcogenide heterostructure. *Nature* **595**, 48–52 (2021).

35. Katsch, F., Selig, M. & Knorr, A. Theory of coherent pump–probe spectroscopy in monolayer transition metal dichalcogenides. *2D Mater.* **7**, 015021 (2019).

**Figures:**

**a)**    **b)**

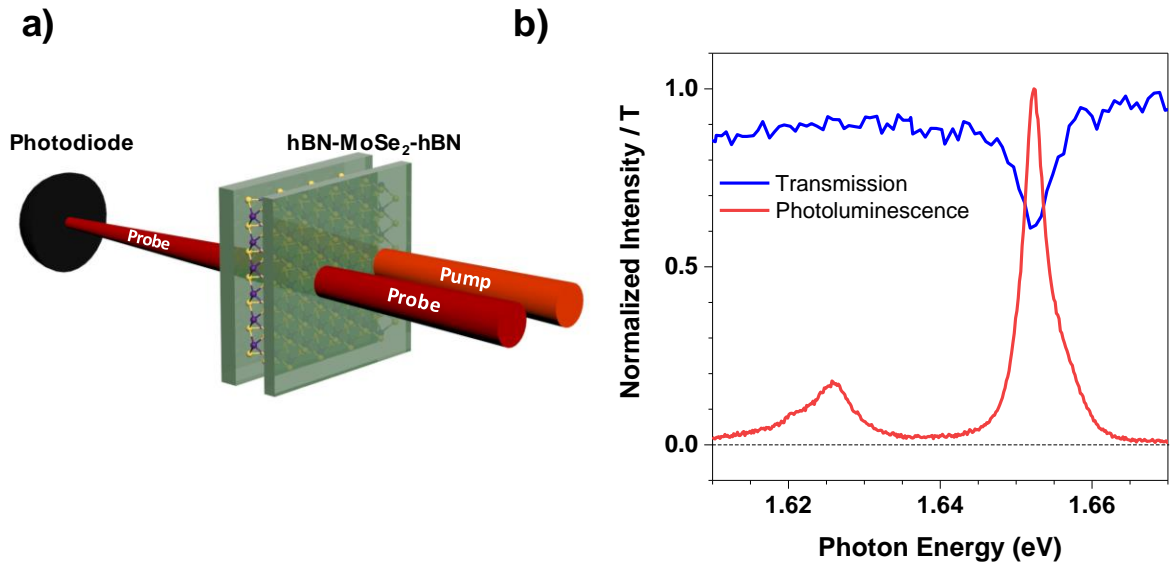

**Figure 1 a)** Cartoon depiction of the steady-state differential transmission experiment on hBN encapsulated $MoSe_2$. **b)** Low temperature (7 K) PL (red) and transmission (blue) spectra of hBN encapsulated $MoSe_2$ sample, showing an exciton energy of 1.652 eV and trion energy of 1.626 eV.

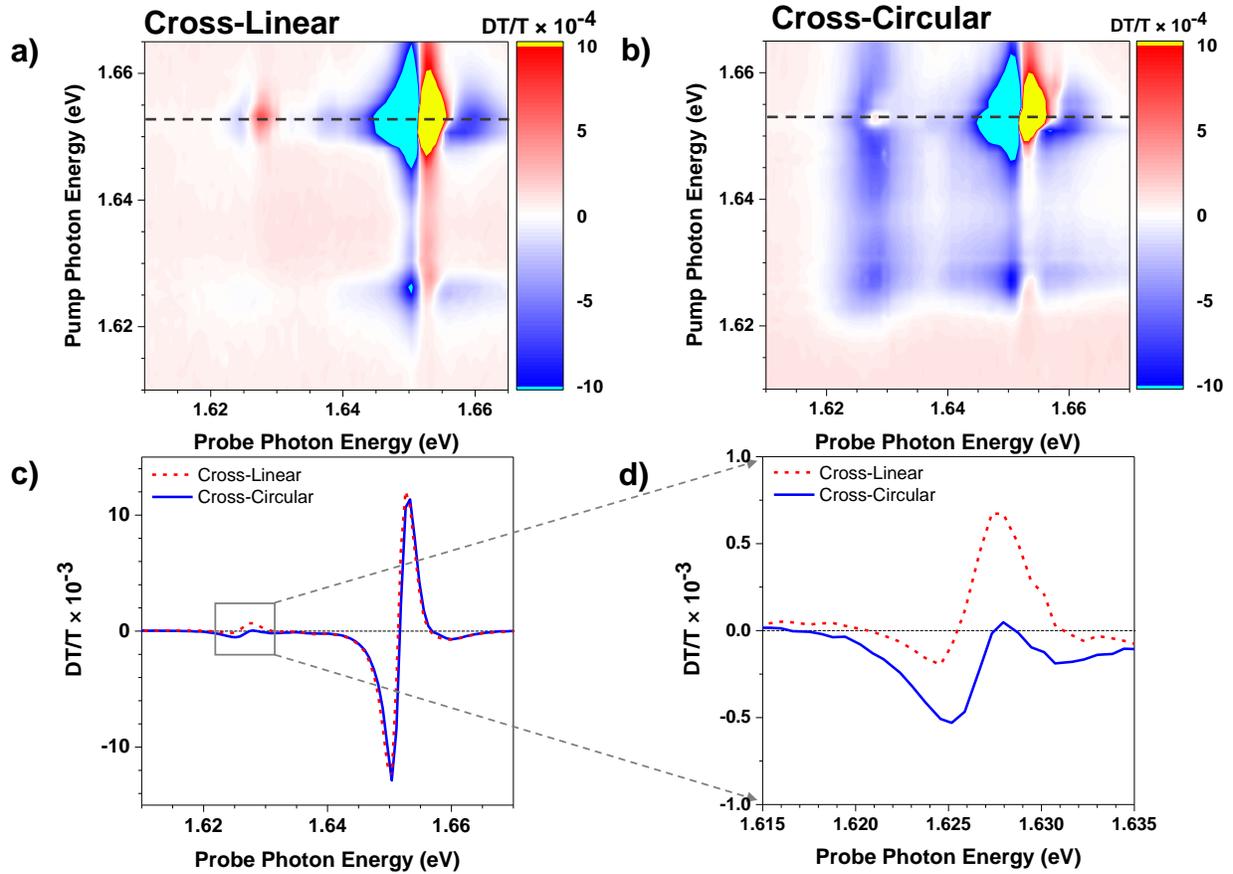

**Figure 2** Experimental DT/T maps on hBN encapsulated MoSe$_2$ as a function of pump and probe photon energies. The probe laser was cross-polarized to reject the pump beam. **a)** DT/T with cross-linearly polarized beams. **b)** DT/T with cross-circularly polarized beams. Yellow and cyan represent saturation of the color scale. **c)** Cross linearly polarized (red) and cross-circularly polarized (blue) line cut taken from the dotted horizontal lines in **a)** and **b),** corresponding to pumping at the exciton energy (1.652 eV). **d)** A zoom-in of the boxed feature in **c)**, near the probe photon energy of 1.626 eV.

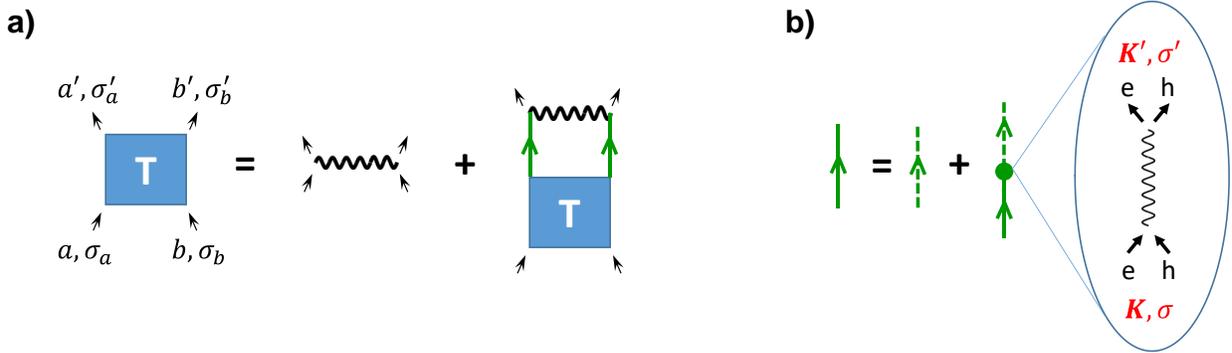

**Figure 3** Schematic visualization of the theoretical model used in the calculation of the nonlinear susceptibility. **a)** T-matrix for 2-particle interaction. The particles are labeled, and each particle can be exciton or trion, treated here as point particles. The corresponding spins are determined by the valley where the exciton (or exciton component in case of trions) is located. The thick wavy line is the 2-particle interaction. The free 2-particle propagator (two green solid lines) depends on the electron-hole (e-h) exchange interaction. **b)** Single-particle propagator (solid green line) renormalized by e-h exchange; the green dashed line is the unrenormalized propagator and the thin wavy line the e-h Coulomb potential. For intravalley exchange, the sums of the electron and hole spins, denoted by $\sigma$, are the same in the valleys K and K'; in intervalley exchange, they are different.

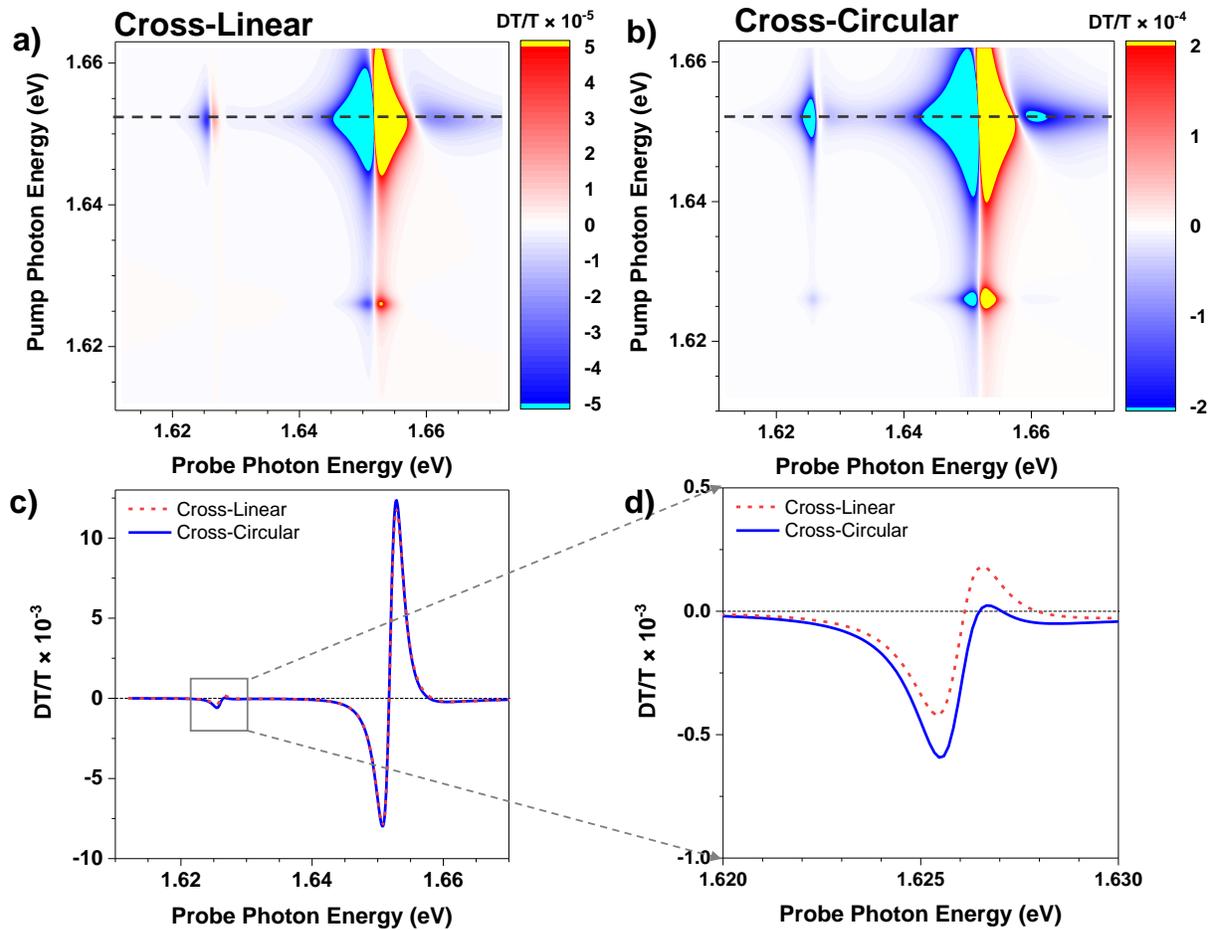

**Figure 4** Theoretical DT/T maps, showing good agreement with experiment in both the **a)** cross-linearly polarized and **b)** the cross-circularly polarized excitation channel. Yellow and cyan represent saturation of the color scale. **c)** Theoretical cross-linearly polarized (red) and cross-circularly polarized (blue) line cuts taken from the dotted horizontal lines in **a)** and **b),** corresponding to pumping at the exciton energy (1.652 eV). **d)** A zoom-in of the boxed feature in **c)**, near the probe photon energy of 1.626 eV.